\documentclass[aps,pra,twocolumn,superscriptaddress,aps]{revtex4}
\usepackage{amsmath,amsfonts,amssymb}

\usepackage{graphicx,graphics}

\begin{document}

\title{Rogue waves in birefringent optical fibers}

\author{Mark J. Ablowitz}
\affiliation{Department of Applied Mathematics, University of
Colorado, 526 UCB, Boulder, CO 80309-0526, USA}

\author{Theodoros P. Horikis}
\affiliation{Department of Mathematics, University of Ioannina, Ioannina 45110, Greece}

\begin{abstract}
Rogue waves in birefringent optical fibers are analyzed within the framework of
the coupled nonlinear Schr\"odinger (CNLS) system. The generation of rogue waves is
frequently associated with modulation instability (MI). It is commonly expected that
since the CNLS equations exhibit larger growth rates they should also produce more rogue
events than their scalar counterparts. This is found to occur only when both equations
are focusing. When at least one component is defocusing, the CNLS
equations may still exhibit larger growth rates, compared to the scalar system, but that does not
necessarily result in more or larger events. The birefringence
angle for which the maximum number of events occurs is also identified and the
nature of the rogue wave is described for the different cases.

\end{abstract}

\maketitle

Due to the history of many large destructive waves in the sea, the phenomena of abnormally large
waves have long been studied in water waves. These waves, generally termed rogue or freak waves,
grow substantially (e.g. a factor of 3 and above) compared to their initial average state
\cite{book}. Remarkably, it has now been demonstrated experimentally \cite{solli} that similar
phenomena occur in fiber optics. These experimental observations were supported by computations on
the scalar nonlinear Schr\"odinger (NLS) equation which is a fundamental model describing optical
phenomena. Indeed the NLS equation plays a crucial role in both water waves and nonlinear optics
\cite{waves}. Thus, two seemingly disparate subjects, and their corresponding phenomena are brought
together through their common description. Motivated by the above research, there have been, now,
numerous studies of rogue waves in optics \cite{erkintalo,dudley,dudley3}.

One of the key underlying features in this subject is the behavior and linearized growth of wave
solutions of the NLS equation via the nonlinear mechanism of self-wave interactions, termed
modulation instability (MI), in optics.  MI is a fundamental property of many nonlinear dispersive
systems and is a well documented and understood phenomenon in optics \cite{trillo6,trillo5,dudley}.
It is a basic process that determines the stability behavior of slowly varying, or modulated waves,
and may initialize the formation of stable entities such as envelope solitons
\cite{mi,agrawal,kivshar_book,trillo7}. Furthermore, it is generally believed that MI is amongst
the several mechanisms leading to rogue wave excitation
\cite{zakharov,zakharov2,akhmediev,onorato_report,baronio3,erkintalo}. Contrary to solitons,
however, rogue waves seem to appear from nowhere, are generally short-lived and the specific
conditions that cause them is still a subject of wide interest.

The scalar NLS system provides information about wave systems with one underlying frequency. But
for several physically relevant contexts this system turns out to be an oversimplified description.
For example, the state of the sea in which rogue waves form is often complex
\cite{onorato,onorato2,horikis} with certain key wave interactions dominating the wave structure.
Similarly in many optical systems, interacting wave components, such as in optical fibers with
randomly varying birefringence, are critical. In some cases the integrable Manakov system
\cite{baronio,baronio2,sara} has been found to be a good model. However, in the general case, the
coupled system that describes complex coupled pulse dynamics in optical fibers is not known to be
integrable making the study of these waves more challenging in terms of analytical results. In
fact, in the description of rogue waves, where the scalar NLS equation is the governing equation,
the Peregrine soliton \cite{peregrine,breather} has been proposed as a model of these events. The
Peregrine soliton is a special type of solitary wave which is formed on top of a modulationally
unstable continuous wave (cw) background; in contrast to other soliton solutions of the NLS it is
written in terms of rational functions. It also has the property of being localized in both time
and space. As such, these solutions may, indeed, be useful to locally describe rogue wave events
\cite{akhmediev2,baronio2}. The nature of these waves will also be discussed later in the text.

Propagation of optical pulses in an elliptically birefringent fiber is governed by a set of coupled
equations that in normalized form are given by \cite{agrawal}
\begin{subequations}
 \begin{gather}
i\frac{{\partial u}}{{\partial z}} + \frac{{{d_1}}}{2}\frac{{{\partial ^2}u}}{{\partial
{t^2}}} + (|u{|^2} +
g|v{|^2})u = 0 \\
i\frac{{\partial v}}{{\partial z}} + \frac{{{d_2}}}{2}\frac{{{\partial ^2}v}}{{\partial
{t^2}}} + (g|u{|^2} +
|v{|^2})v = 0
\end{gather}
\label{cnls}
\end{subequations}
where the parameter $g$ is defined by the ellipticity angle $\theta$ as $\displaystyle
g=\frac{2+2\sin^2\theta}{2+\cos^2\theta}$. Its values vary from $2/3$ $(\theta=0^\circ)$, in the
case of a linearly birefringent fiber, to $2$ ($\theta=90^\circ$), for a circularly birefringent
fiber. The dispersion values are denoted by $d_1$ and $d_2$. We will consider three different cases
in this article. The numerical values of the dispersions are always $1$ but we vary the signs such
that: $d_1=d_2=1$ in the focusing case, $d_1=d_2=-1$ in the defocusing case and for completeness
$d_1=-d_2=-1$; hereafter, the latter case is termed the semi-focusing regime. While this case does
not occur in this particular application we include it as a means to explain interesting phenomena
which can occur in other applications. As is standard, $u(t,z)$ and $v(t,z)$ are the complex
amplitudes in each polarization mode.

The fundamental cw solution of \eqref{cnls}, is
\[
u=u_0 e^{i ( u_0^2+g v_0^2) z},\quad v=v_0 e^{i ( g u_0^2+ v_0^2) z}
\]
where $u_0$ and $v_0$ are real constants. Now consider small perturbations to these cw solutions,
so that,
\[
u(t,z)=[u_0 + u_1(t,z)]e^{i \theta_1 z},\,
v(t,z)=[v_0 + v_1(t,z)]e^{i \theta_2 z}
\]
where $\theta_1=u_0^2+g v_0^2$ and $\theta_2=g u_0^2+ v_0^2$. The small perturbations are assumed
to behave as $\exp[i(kt-\omega z)]$ and these small-amplitude linear waves obey the following
dispersion relation:
\begin{gather}
{\omega ^4} - {k^2}[(d_1^2 + d_2^2){k^2} - 2({d_1}u_0^2 + {d_2}v_0^2)]{\omega
^2}\nonumber\\
+ {d_1}{d_2}{k^4}[{d_1}{d_2}{k^4} - 2({d_2}u_0^2 + {d_1}v_0^2){k^2} + 4(1 -
{g^2})u_0^2v_0^2] = 0.
\label{dispersion}
\end{gather}
This equation is a bi-quadratic and can be solved explicitly for $\omega=\omega(k)$. It is,
therefore, relatively straightforward to establish when $\omega$ is real/complex, i.e. when the
system is modulationally stable/unstable \cite{agrawal}.

\begin{figure}[ht]
\centering
\includegraphics[scale=.35]{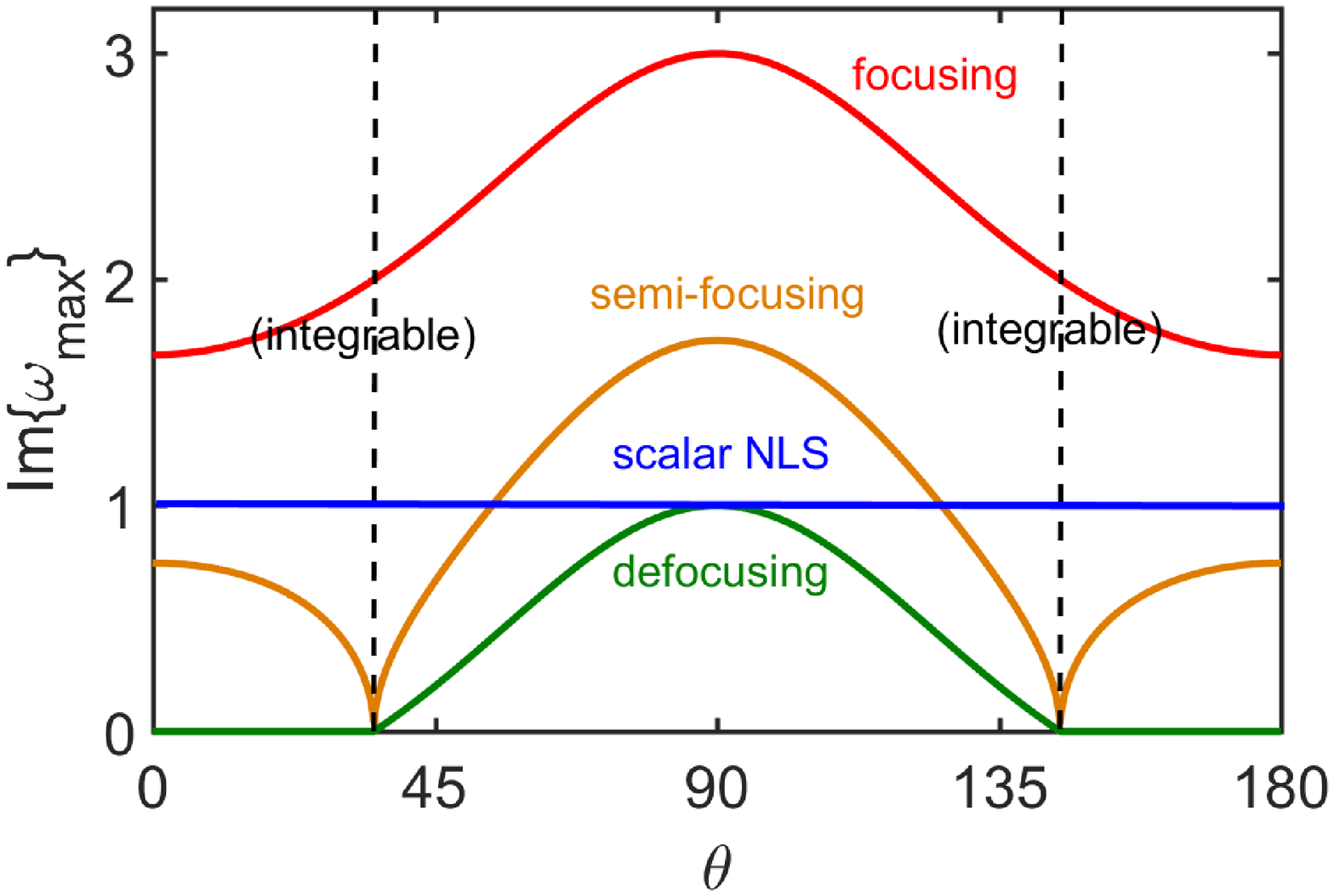}\\
\includegraphics[scale=.35]{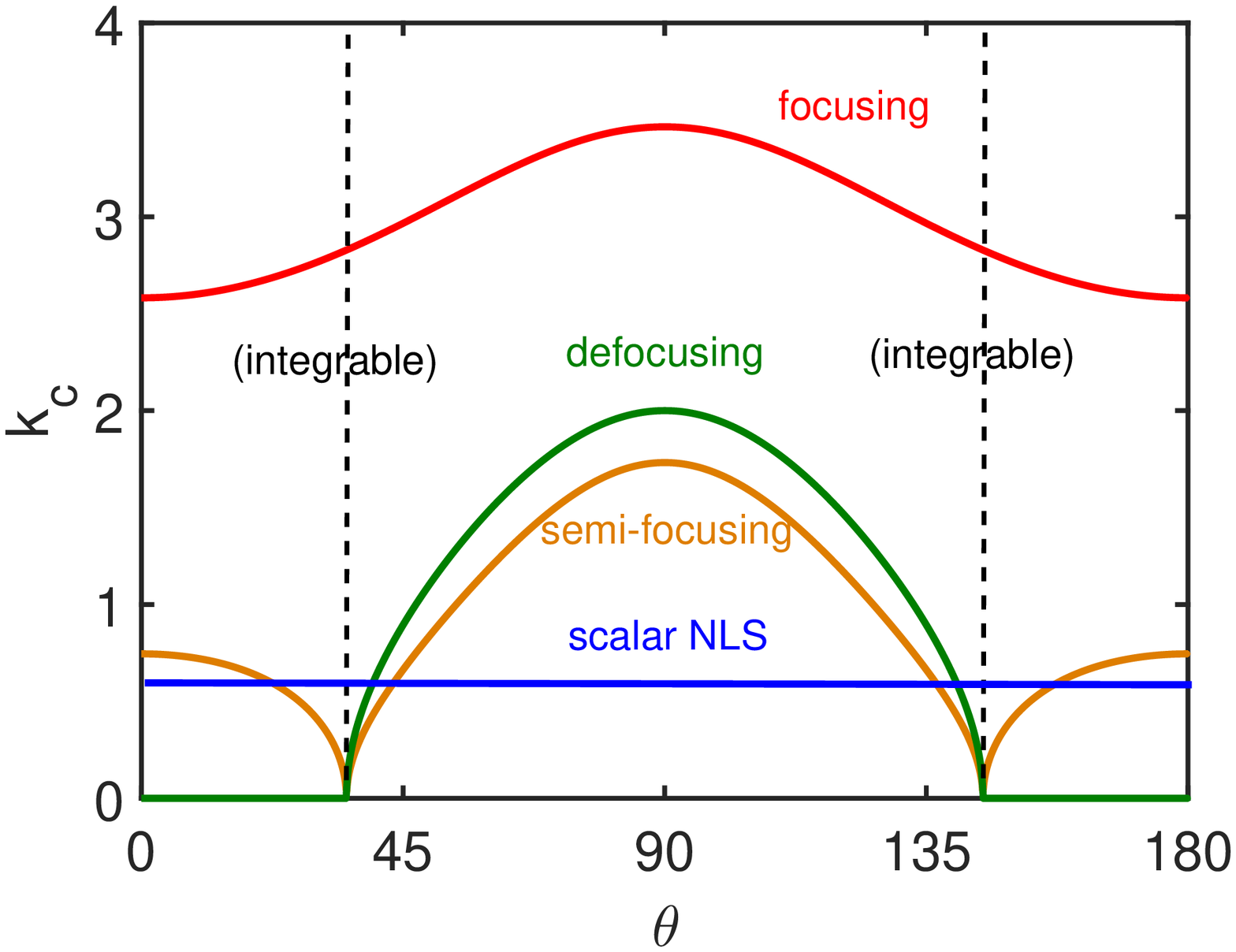}
\caption{(Color Online) Top: The maximum growth rate vs. the ellipticity
angle $\theta$. Bottom: The critical wave number, $k_c,$ vs. the ellipticity
angle $\theta$. The dashed vertical lines identify the integrable $(g=1)$ case.}
\label{growth}
\end{figure}

Furthermore, by solving \eqref{dispersion} one can identify three critical values: the location of
the maximum growth rate, $k_{\mathrm{max}}$, which in turn corresponds to the maximum growth rate,
$\mathrm{Im}\{\omega_{\mathrm{max}}\}$, and a wavenumber, $k_c$, which characterizes the length of
the instability band. The maximum growth rate characterizes the time needed for MI to be exhibited
(the higher $\mathrm{Im}\{\omega_{\mathrm{max}}\}$ the sooner MI occurs), while $k_c$ is the
maximum value of the wave numbers, centered around $k_{\mathrm{max}}$, for which the system is
unstable. These two critical values, for \eqref{cnls}, are shown in Fig. \ref{growth} as functions
of the angle $\theta$. Hereafter, $u_0=v_0=1$.

Interestingly, both values are maximized for circular birefringence, i.e. for $\theta=90^\circ$ and
take their minimum values for linear birefringence $\theta=0^\circ$. This means that for
$\theta=90^\circ$  the instability will occur sooner and there also more wavenumbers for which the
system is unstable. Furthermore, as seen from these figures, the coupled system differs
significantly from its scalar counterpart. Indeed, in the focusing case the MI growth rate is
larger than the corresponding scalar NLS equation, while in the defocusing case the system can be
modulationally unstable (for some angles) with growth rates comparable to its scalar focusing
analogue. What is also important for our study, is that, in general, the coupled semi-focusing
system exhibits higher growth rates and the range of unstable wavenumbers is also significantly
wider than the scalar focusing system. That, typically, would suggest that the coupled system
should exhibit more rogue events as it is more unstable.

To test this, we integrate numerically \eqref{cnls} using a pseudospectral method in space and
exponential Runge-Kutta for the evolution \cite{kassam} in a computational domain $t
\in[-100,100]$, $z\in[0,20]$. An appropriate initial condition would be a wide gaussian of the form
\[
u(t,0)= v(t,0)=e^{-t^2/2 \sigma^2},\; \sigma=30
\]
perturbed with 10\% random noise; random noise ensures that initially the two conditions will
always be different so that the system would not degenerate to the case $u\equiv v$. A wide
gaussian with randomness added is a prototype of a set of broad/randomly generated states which can
potentially excite more that one wave numbers as it can be regarded as a Fourier series of
different cw waves of different $k$'s. For each polarization angle $\theta$ we perform $10^5$
trials. In each trial we measure the highest wave amplitude and introduce the quantity
\[
\eta(t,z)  = \sqrt {\frac{{|u(t,z){|^2} + |v(t,z){|^2}}}{\max\{|u(t,0){|^2} +
|v(t,0){|^2}\}}}
\]
which is the ratio of the highest combined wave amplitude (complex electromagnetic field) to the
maximum of the initial condition. This measures the relative growth in amplitude from an initial
state. Here we consider a rogue event as one in which $\eta(t,z)$ at some value of $z$ is at least
three times its maximum  initial value.

\begin{figure}[ht]
\centering
\includegraphics[scale=.2]{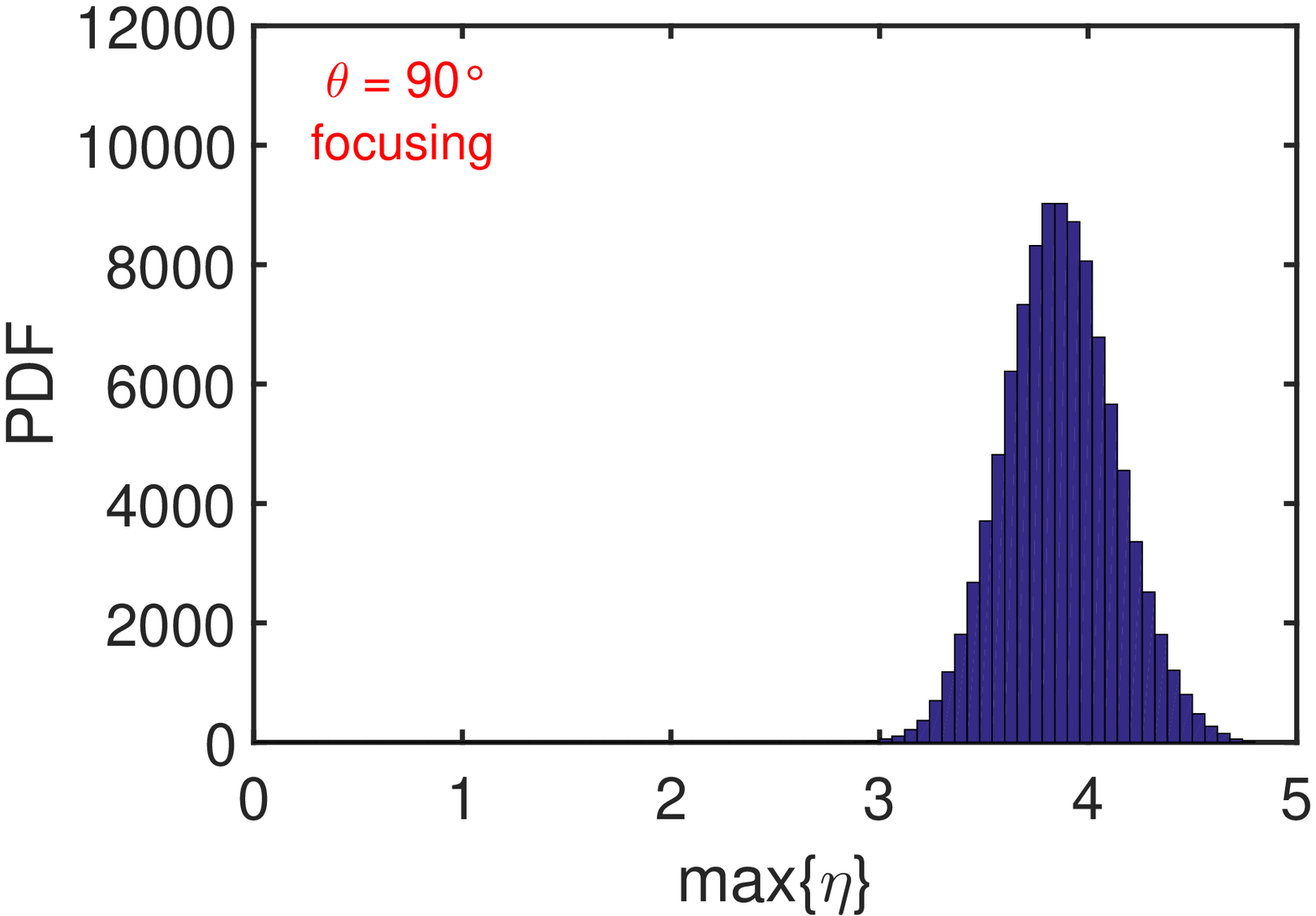}
\includegraphics[scale=.2]{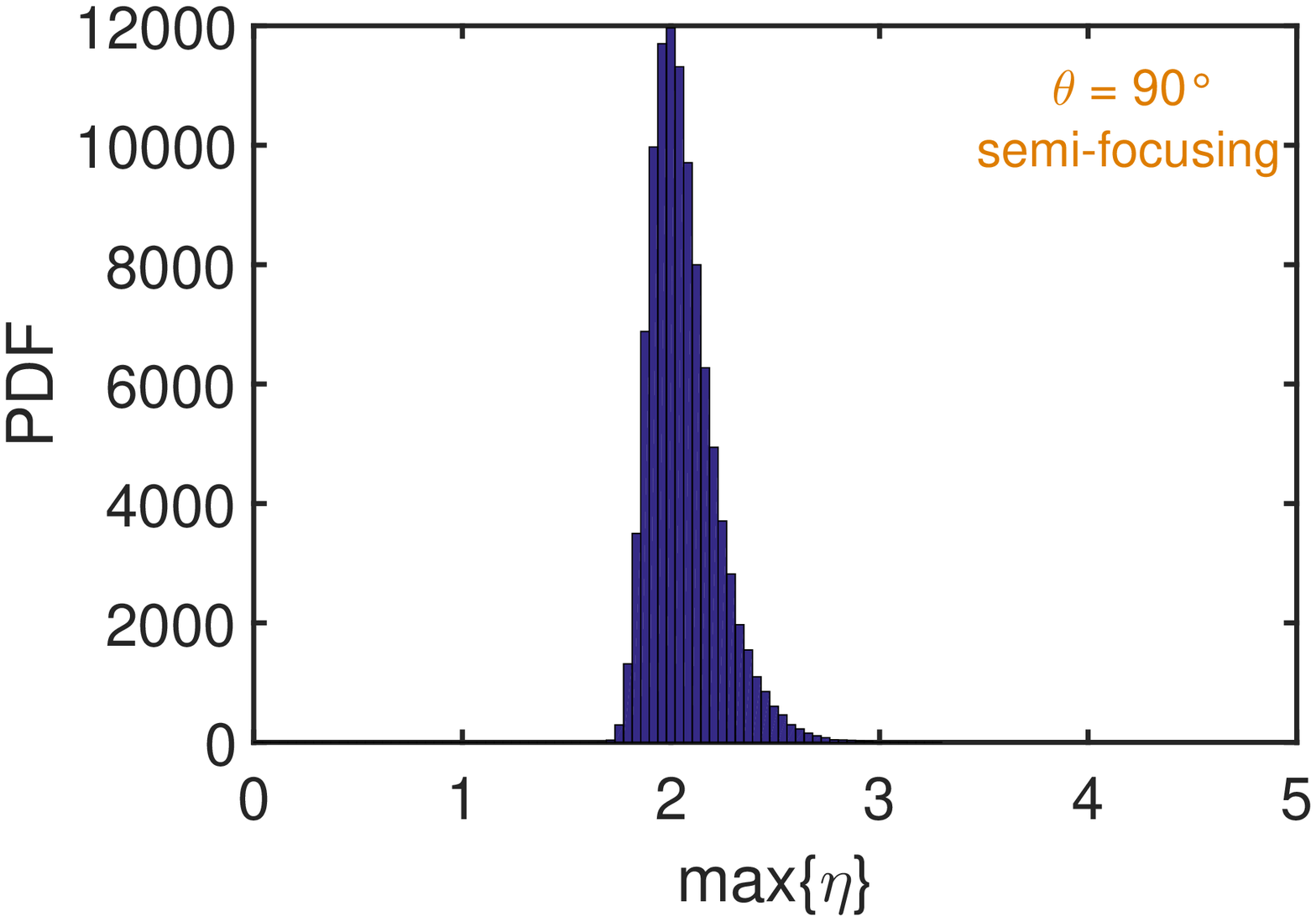}\\
\includegraphics[scale=.2]{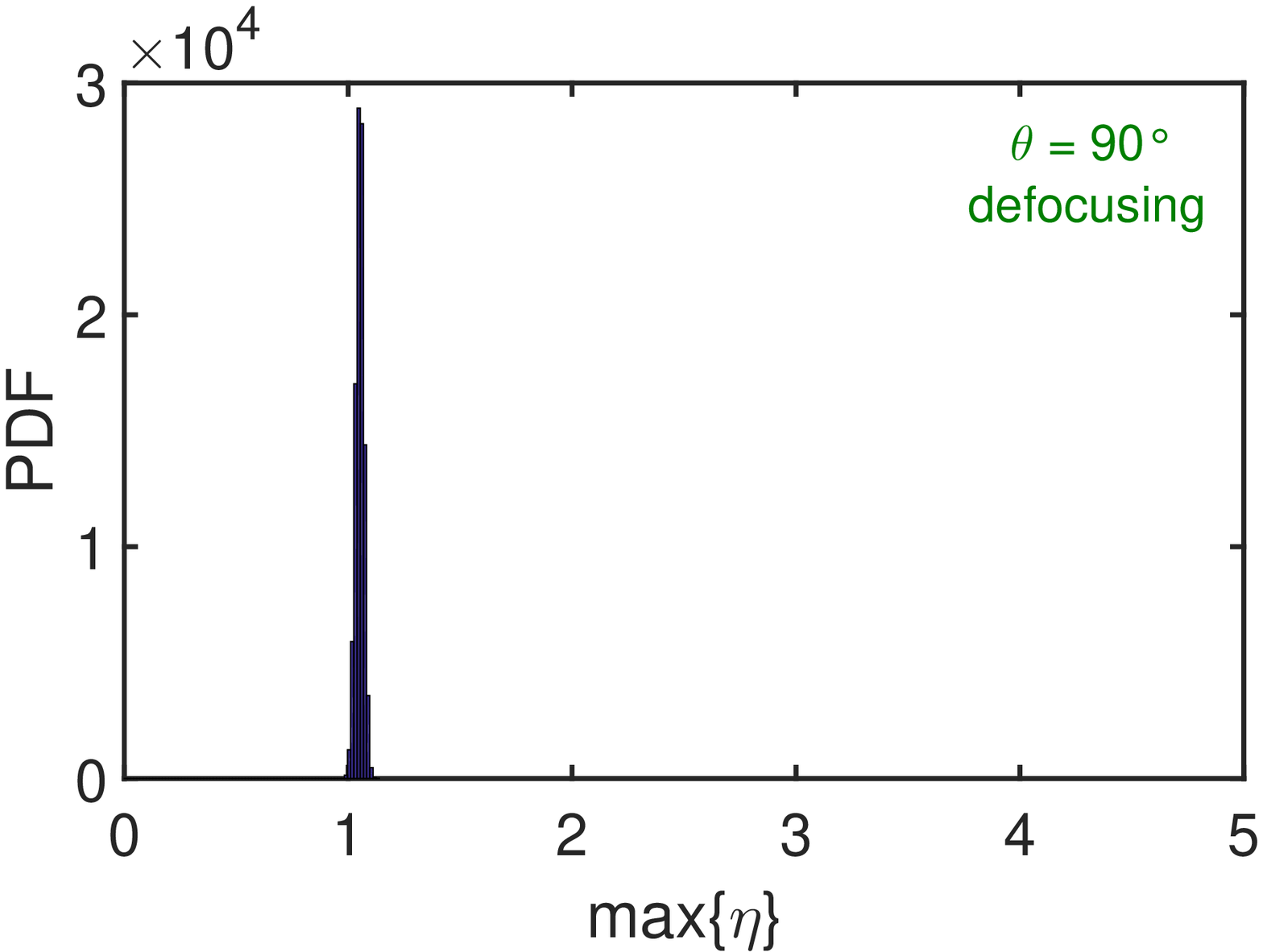}
\includegraphics[scale=.2]{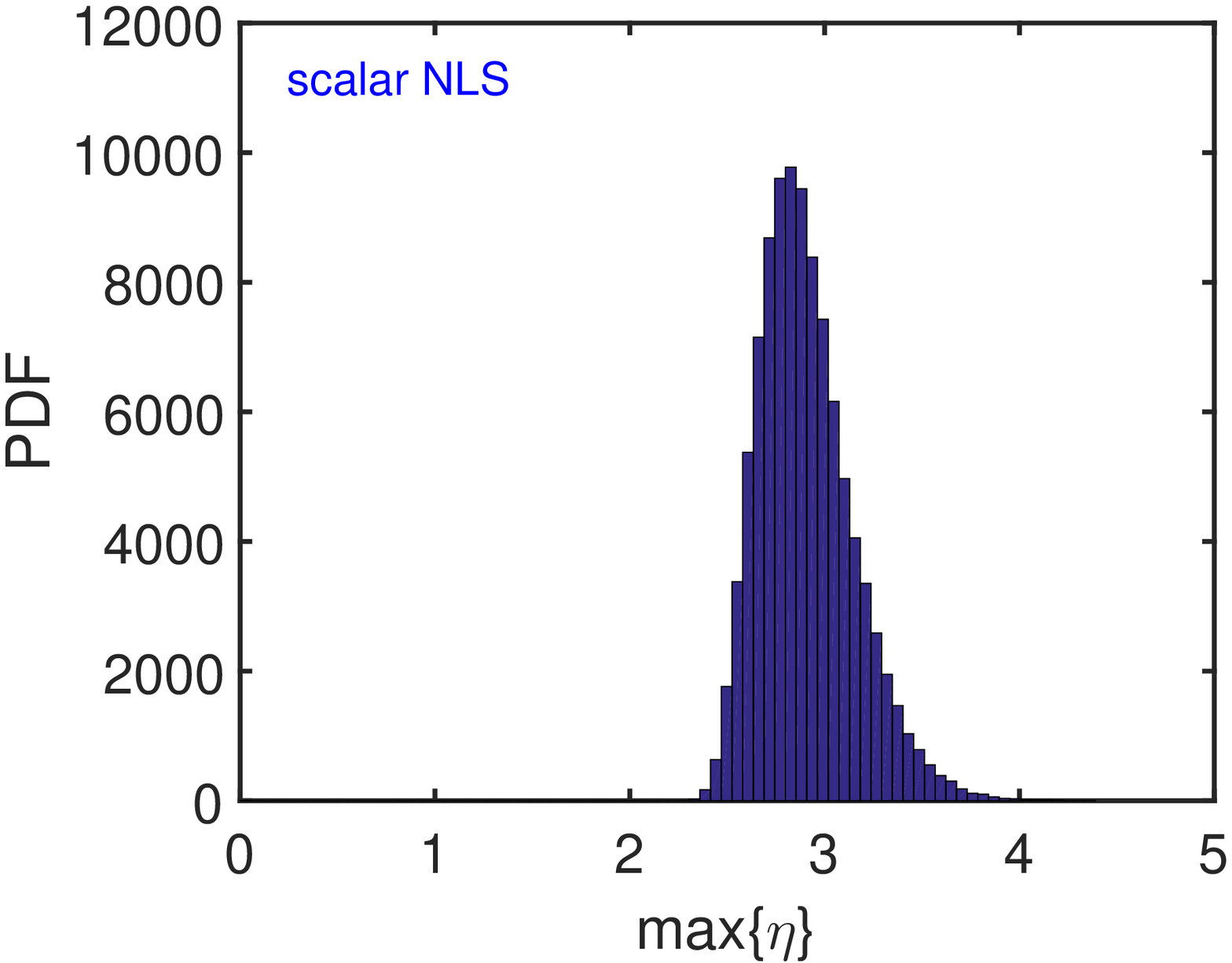}
\caption{(Color Online) The PDFs for the highest growth rate angle, $\theta=90^\circ$,
and the corresponding scalar system.}
\label{pdf}
\end{figure}

We show in Fig. \ref{pdf}, for $\theta=90^\circ$ where all critical values are maximized, the
probability density functions (PDFs) for all three cases and the results for the scalar NLS
equation.

These PDFs indicate that for the focusing system at $\theta=90^\circ$ the mean value of the
distribution has a significant shift towards larger values, as compared to the scalar NLS equation.
On the other hand the results from the semi-focusing and defocusing cases do not show a similar
shift under the same comparison. Indeed, in the semi-focusing case the maximum growth rate is
higher than the scalar system while in the defocusing case  they are equal. However, in the
semi-focusing case the mean is shifted towards smaller values as compared to the scalar case, and
in the defocusing case there are no rogue events.

\begin{figure}[ht]
\centering
\includegraphics[scale=.35]{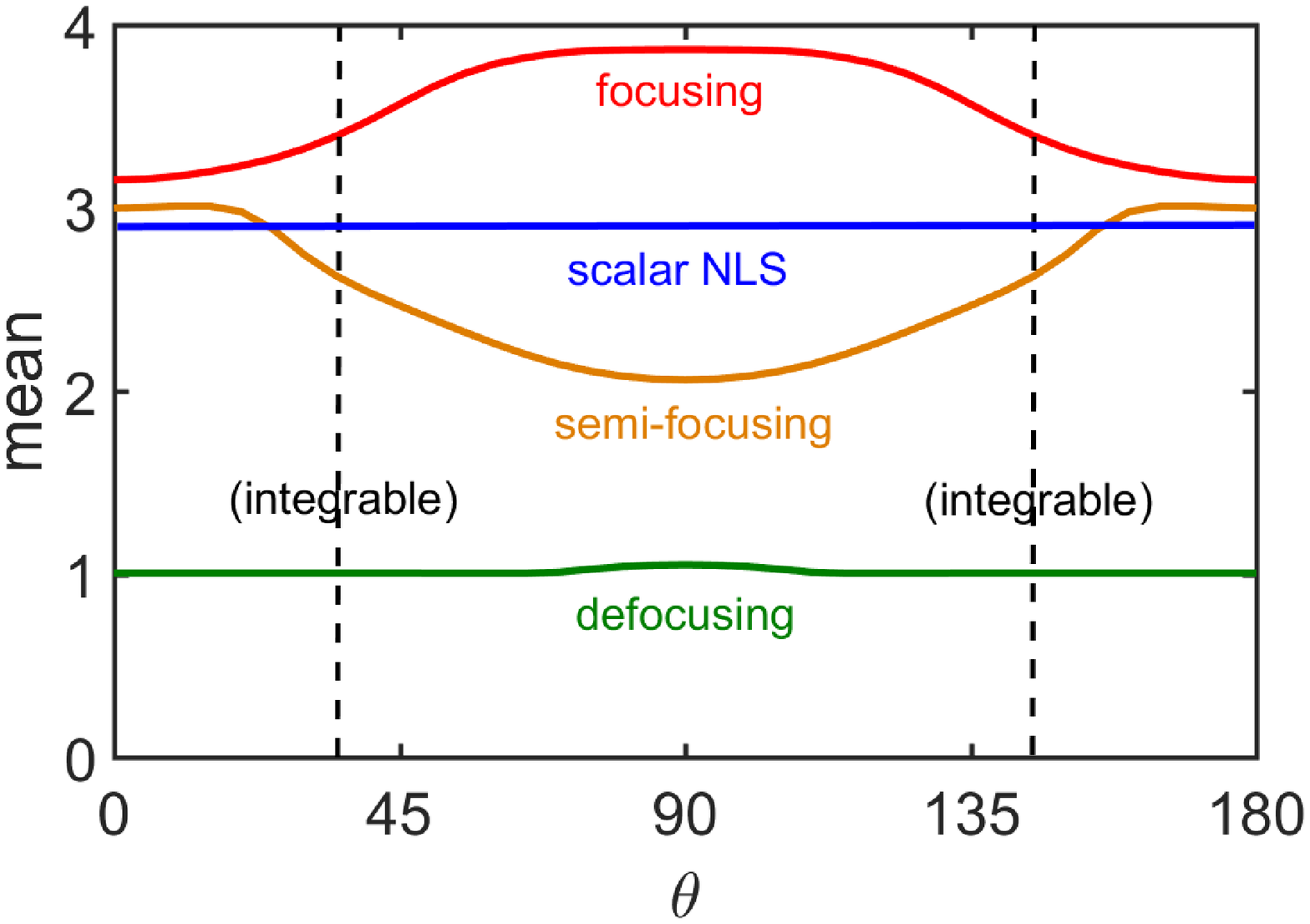}\\
\includegraphics[scale=.35]{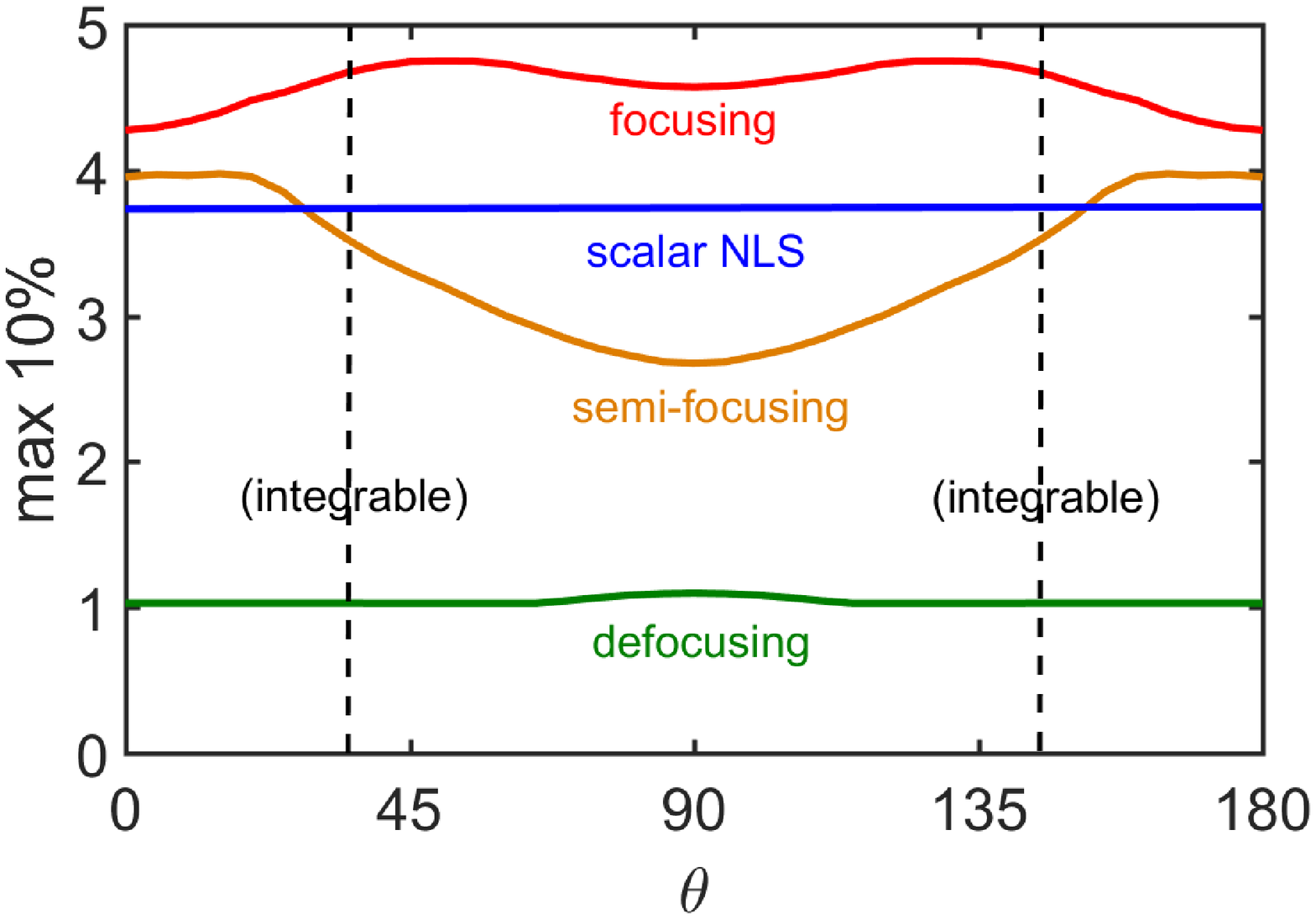}
\caption{(Color Online)  The mean value of the PDFs (top) and the mean value of the max
10\%
events
(bottom) with the the ellipticity angle $\theta$.
The dashed lines correspond to the integrable $(g=1)$ case.}
\label{max.mean}
\end{figure}

To further investigate the above observation, we perform the same analysis for different values of
$\theta$. In Fig. \ref{max.mean} we depict the change of the mean value of rogue events with
$\theta$ as well as the change of the top $10$\% of the highest valued events. As clearly seen,
rogue events do not follow the common belief that they are intimately related to the MI growth
rates.

There are essentially three distinct cases here, which are not distinguished by the relative MI
analysis. The three cases refer to the three different systems: focusing, semi-focusing and
defocusing. In the first case, there is a significant increase in the number and severity of rogue
events; this is consistent with the increase in the growth rates and size of the instability
region. However, in the approximate region $60^\circ<\theta<90^\circ$, the mean remains essentially
stable. The picture is different with regard to the maximum events where there is a slight
decrease. Overall, however, in the pure focusing case there is a significant increase in both the
number and severity of events. On the other hand, in the defocusing case, the number of rogue
events is essentially negligible.

This contradicts the observation that for $35^\circ<\theta<90^\circ$ the system is unstable with
nontrivial  growth rates and in turn the belief that this should result into more rogue events. In
the semi-focusing case, while the growth rates are increasing for these values of $\theta$ both the
mean and max values are decreasing. In fact, the system exhibits a minimum mean and max values for
$\theta=90^\circ$ while both $\mathrm{Im}\{\omega_{\mathrm{max}}\}$ and $k_c$ are maximized.
Furthermore, the scalar NLS equation exhibits larger mean values and maxima for a wide range of
angles. These observations clearly contradict the belief that higher MI growth rates should lead to
more numerous rogue waves; these results are in agreement with the observations of Ref.
\cite{horikis} in deep water waves.

\begin{figure}[ht]
\centering
\includegraphics[scale=.3]{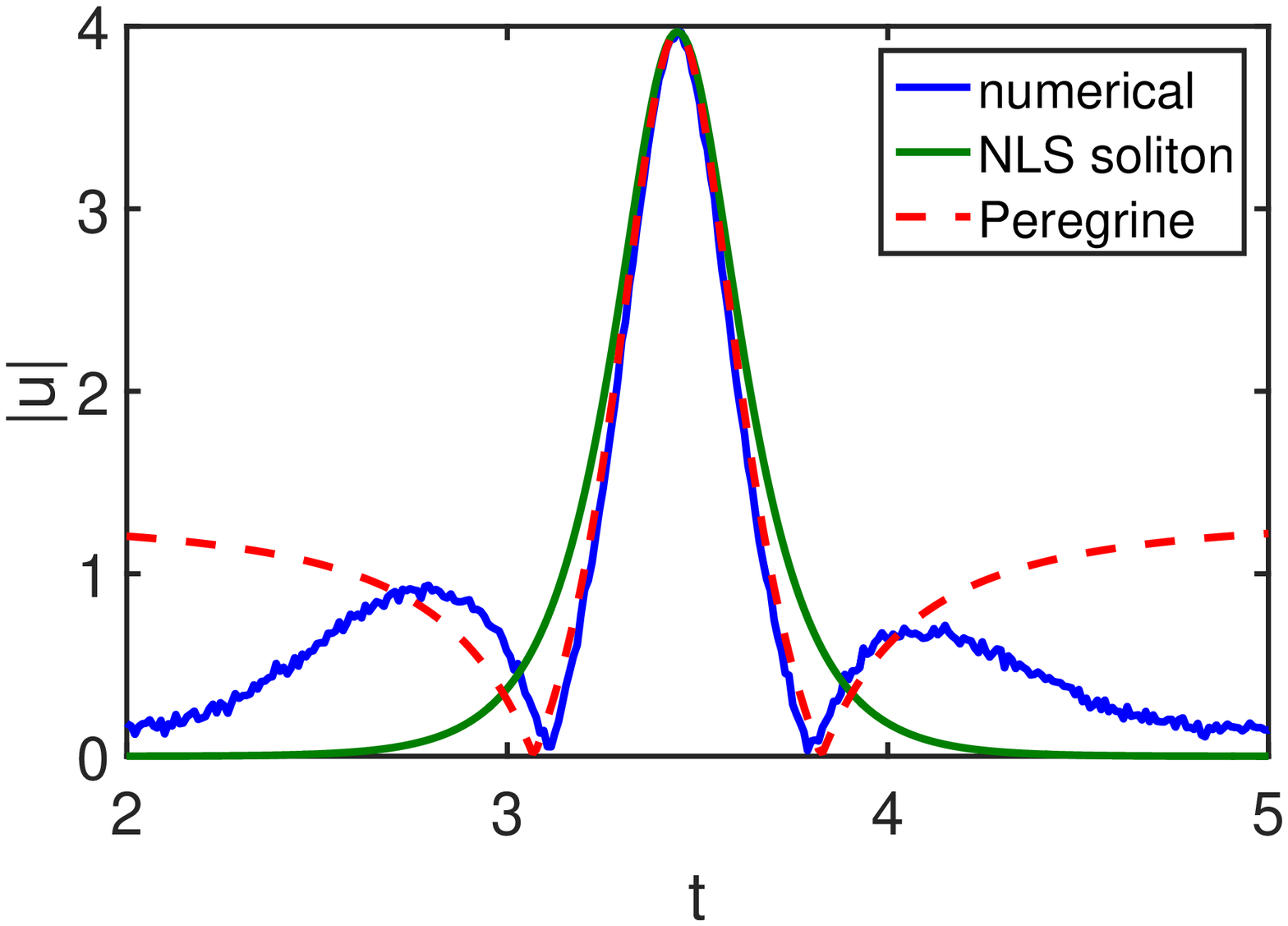}\\
\includegraphics[scale=.3]{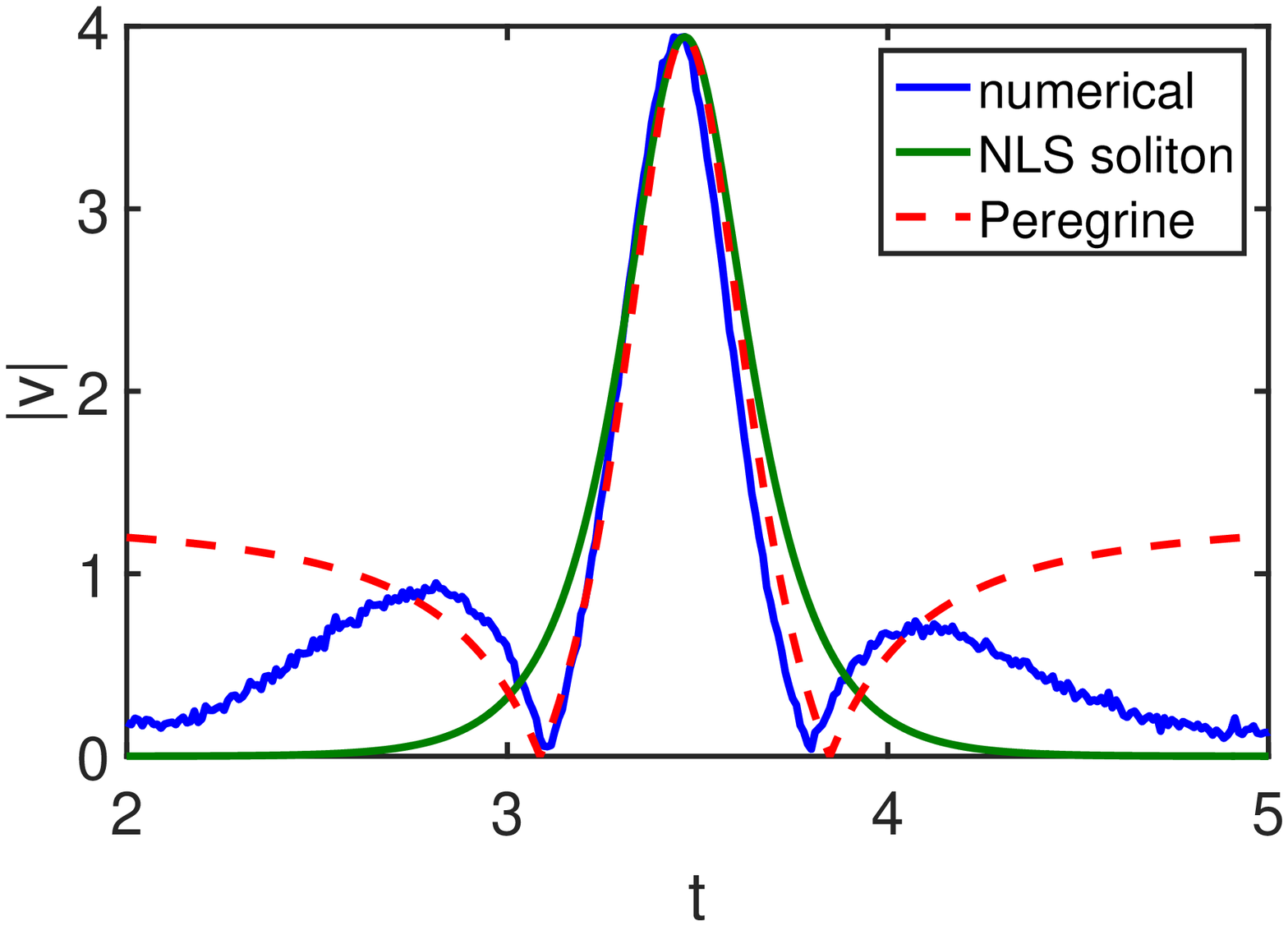}
\caption{(Color Online) The maximum wave amplitude for the focusing case ($\theta=90^\circ$). The numerical
observation is fitted against the Peregrine and single soliton solutions of \eqref{nls}.}
\label{rogue_focusing}
\end{figure}

Lastly, we exploit the nature of these rogue events. This also helps provide additional context to
the above observations. We focus only on the two focusing and semi-focusing cases where rogue waves
are found. We fix the ellipticity angle to $\theta=90^\circ$ and zoom in around a typical maximum
event in both components of the field $u$ and $v$. We begin with the focusing case, in Fig.
\ref{rogue_focusing}.

In this case, we observe that a very good fit to a typical rogue event is described by the scalar
NLS equation
\begin{equation}
i\frac{{\partial u}}{{\partial z}} + \frac{{{d}}}{2}\frac{{{\partial ^2}u}}{{\partial
{t^2}}} + (g+1)|u|^2u =
0
\label{nls}
\end{equation}
where $d_1=d_2=d$ and $u\equiv v$. This equation admits the so-called Peregrine soliton
\cite{peregrine}
\begin{equation}
u(t,z) = {u_0}\left[ {1 -
\frac{{2d[1 + 2i(g + 1)u_0^2z]}}{{d/2 + 2d{{(g + 1)}^2}u_0^2{z^2} + 2(g + 1)u_0^2{t^2}}}}
\right]{e^{i(gu_0^2)z}}
\label{rational}
\end{equation}
while the corresponding single soliton solution reads
\begin{equation}
u(t,z) = {u_0}{\rm sech}\left( \sqrt {(g + 1)/d} {u_0}t \right){e^{i[(g + 1)u_0^2/2]z}}.
\label{soliton}
\end{equation}
The free parameter $u_0$ represents the amplitude of the wave in both cases.

\begin{figure}[ht]
\centering
\includegraphics[scale=.3]{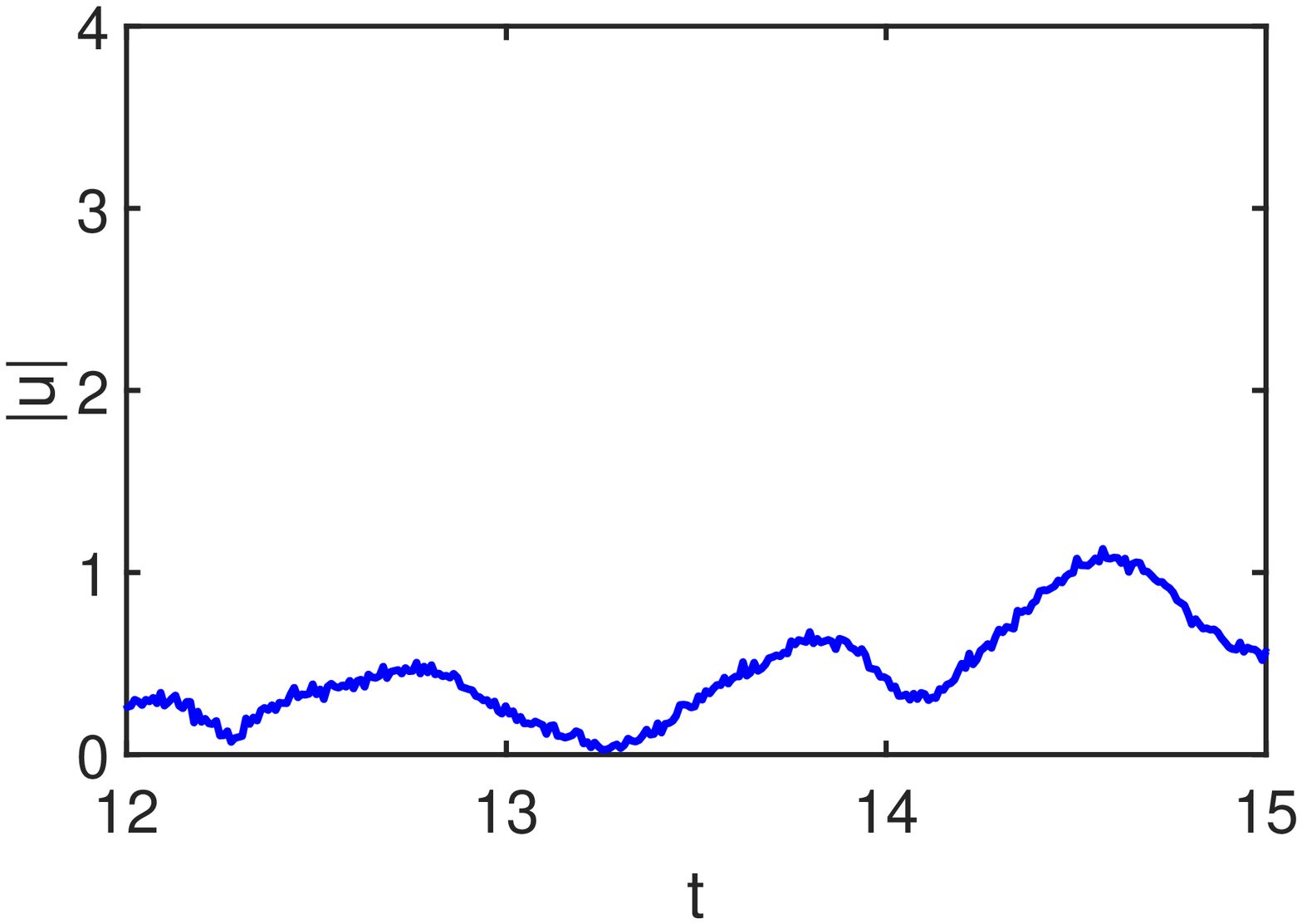}\\
\includegraphics[scale=.3]{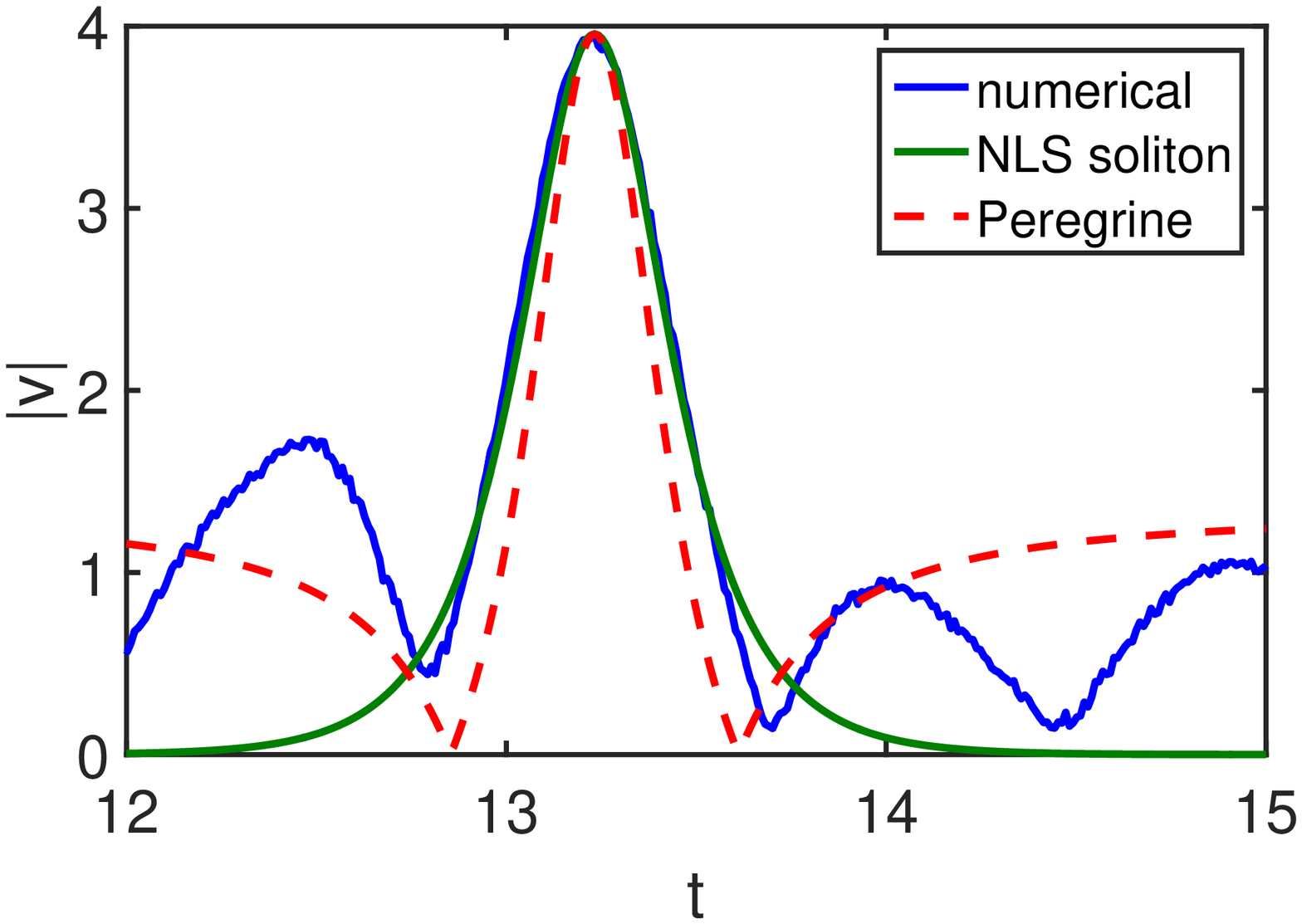}
\caption{(Color Online) The maximum wave amplitude for the semi-focusing case ($\theta=90^\circ$).
The numerical observation is fitted against the Peregrine and single soliton solutions
of \eqref{nls}.}
\label{rogue_semi}
\end{figure}

This situation can occur because the reduction $u=v$ is possible in this case as \eqref{cnls} are
symmetric. On the other hand when the system is not symmetric, $d_1=-d_2$, the computations
indicate that the first component ($u$) does not contribute significantly to the rogue wave
formation and the rogue event is well approximated by the solitonic solution \eqref{soliton}. We
see this exemplified in Fig. \ref{rogue_semi}.

Furthermore the largest rogue events seem to occur when both equations are unstable, i.e. when both
contribute to the formation of extreme events. In this case, the rogue wave is well approximated
(for the symmetric system) by the rational solution, \eqref{rational}. On the other hand, when
symmetry is lost and only one equation dominates the rogue wave formation, the resulting wave is a
soliton of the respective  unstable scalar equation, \eqref{soliton}. This is also consistent with
the findings of Ref. \cite{horikis}, where in the unstable case the rogue waves were well
approximated by hyperbolic secants; they did not exhibit algebraic decay.

To conclude, we have studied the occurrence of rogue waves in birefringent optical fibers. For
certain parameter regimes, in the focusing case, e.g. in the circularly birefringent case
($\theta=90^\circ$),  a major increase in the number of rogue events results from the CNLS system
as compared to the scalar NLS equation; this correlates with an increase in the growth rate and
size of the MI region. But larger MI growth rates do not always lead to more rogue events: i.e. in
the defocusing and semi-focusing cases. Furthermore, we find that the Peregrine soliton is only a
good approximation of the focusing CNLS system when the rogue event in both components are nearly
equal and, in turn, the system is approximated by the reduced scalar NLS equation. In the
semi-focusing case rogue events are such that the defocusing component is negligible and the
semi-focusing CNLS system reduces to the scalar focusing NLS equation;  here the event is well
approximated by the one soliton solution of the relative NLS equation.

MJA is partially supported by NSF under Grant DMS-1310200.

\end{document}